\documentstyle[12pt,myown,twoside,axodraw]{report}
\begin{document}
\protect{\pagenumbering{arabic}}
\protect{\setcounter{page}{1}}
\begin{center}
{\Large{\bf How Is Nature Asymmetric ?}}\\ \vspace{0.5 cm}
{\large{\sf 1. Discrete Symmetries in Particle Physics and Parity Violation}}\\
\vspace{0.5 cm}
%{\sf 1. Background and Parity Violation}\\
%\vspace{0.5 cm}
B. Ananthanarayan, J. Meeraa, Bharti Sharma, \\
Seema Sharma and Ritesh K. Singh\\
\end{center}
{\bf This two-part article considers certain
fundamental symmetries of nature, namely the discrete symmetries of parity (P),
charge conjugation (C) and time reversal (T), and their possible
violation. Recent experimental results are discussed in some depth.
In the first part of this article we present a general background
and discuss parity violation.}\\\\
{\large{\bf Introduction}}\\\\
In day-to-day life when we use the word {\it symmetry}, we mean a geometric
property of an object by virtue of which it remains unchanged on performing
some transformation on it. For example, a square looks the same when rotated by
$90^{\circ}$, a circle remains unchanged when rotated by any angle about its
center, a regular hexagon looks indistinguishable when rotated by multiples of
$60^{\circ}$.\\\\
In physics, if equations of motion for a given physical system remain unchanged
after some kind of transformation on the system, then it is said to be
symmetric on the basis of its physical behaviour which is unaffected by these
transformations. Furthermore, we need to know the symmetries appropriate to a
physical system because all conservation laws in physics are consequences of
some underlying symmetries. For example, the law of conservation of linear
momentum is a consequence of the {\it homogeneity} of space, i.e., {\it
translational symmetry} of space. This is possible only in the absence of
external forces. If there is an external force then the point at which it acts
is a special point and it is not physically identical with other points. Thus
homogeneity of space ensures the absence of external forces and, along with the
law of inertia, establishes the law of conservation of linear momentum.\\\\
Knowledge of conservation laws is very important for studying any system or
solving physical problems. Hence the study of symmetries of the system is
equally important. We define symmetry thus : \\\\
{\it If a physical system undergoes certain transformations (e.g.,
translations, rotations), and if the transformed system looks identical (on the
basis of physical observables) to the untransformed one, then those
transformations will be called symmetry transformations and the system will be
said to posses those symmetries.}\\\\
Symmetries are of two types : continuous and discrete. Continuous symmetry
transformations are labelled by parameters each of which can take any value in
a given range. On the other hand, discrete symmetry transformations are
labelled by a set of integers or discrete numbers. The translational symmetry
of empty space is a continuous symmetry because translation by any amount
leaves the space unchanged. Rotation of a circle in its plane is also a
continuous symmetry. Rotation of a square is a discrete symmetry as only
rotations by $n \times 90^{\circ}$, for all integer $n$, leaves the square
unchanged.\\\\
In this article we shall be concerned with symmetries of the laws that describe
and govern the interactions of {\it elementary particle physics} (see Box 1).
We will be
dealing with the implications of certain {\it bilateral} symmetries (see next
section) to these interactions and their possible violations.\\\\
{\large{\bf Bilateral Symmetries}}\\\\
Those discrete symmetry transformations whose double operation is the same as
no operation are called bilateral symmetries. For example, mirror reflection is
a bilateral transformation since mirror reflection of a mirror reflection is as
good as no reflection. In 3-dimensions, if the mirror is kept parallel to $xy$-
plane, then the reflection will reverse the direction of $z$ axis while
reflection of this reflected image brings the $z$-axis back to its initial
configuration. If we denote the bilateral transformation operator which acts on
the system $S$ by $B$ then we have
$$B(  B \ S  ) \ = \ S$$
which means for all bilateral operators, $B^2$= Identity.\\\\
In the following we will talk about three kinds of bilateral symmetries namely,
parity or spatial inversion (P), charge conjugation (C) and time reversal (T).
\\\\
{\bf (a) Parity - the left-right symmetry}\\\\
Parity transformation refers to the inversion of spatial co-ordinates with
respect to the origin, i.e.,
$$x \longrightarrow -x,\hspace{0.5cm}y \longrightarrow -y,
\hspace{0.5cm}z\longrightarrow -z$$
which is the same as $\vec{\bf r}\rightarrow -\vec{\bf r}$. From this
definition, it is clear that parity is a bilateral transformation. In 3-
dimensions, parity is also called left-right symmetry. If you align the thumb,
fore finger and the middle finger of your right hand perpendicular to each
other with thumb and fore finger in the plane of your palm, then this describes
a right-handed co-ordinate system with $x$, $y$ and $z$ axes pointing along
thumb, forefinger and middle finger respectively. After inversion of co-
ordinates, if you try to align your thumb and fore finger with $x$ and $y$
axes, then the middle finger will point opposite to $z$ axis. But if you try
the same with your left hand, you will find that all the three axes match with
the fingers. This co-ordinate system is left-handed. Thus, parity being a good
symmetry of a physical system implies that the system is left-right symmetric.
\\\\
Under parity operation $P$, a function $f({\bf r})$ transforms to  $f({\bf -
r})$, and if these two functions are the same upto a sign then the function
$f({\bf r})$ will be said to have definite parity. For example,
$$\cos(x)\stackrel{P}{\longrightarrow}\cos(-x) \ = \ \cos(x)\hspace{1cm} {\rm
even \ parity}$$
$$\sin(x)\stackrel{P}{\longrightarrow}\sin(-x) \ = \ -\sin(x)\hspace{1cm} {\rm
odd \ parity}$$
If  $f({\bf -r})$ and  $f({\bf r})$ are of different forms then the function
deos not have definite parity. The solutions of parity symmetric physical
equations may be of definite parity. In the context of elementary particle
physics it becomes necessary to assign such parity {\it quantum numbers} to
particle states (wave functions). The parity of the wave-function of single
particle state is called its {\it intrinsic parity}.\\\\
{\bf (b) Charge Conjugation - $C$}\\\\
Charge conjugation reverses the sign of electric charge of a particle along
with all of its other internal quantum numbers, such as, strangeness, baryon
number, lepton number, and leaves all other quantum numbers unchanged. Symmetry
under $C$ means that interaction of two particles is independent of the sign of
their internal quantum numbers and charge. In other words, this symmetry
implies interaction of two particles is exactly identical to interaction of the
corresponding {\it anti-particles}, where we define anti-particles as the
charge conjugate counterpart of the corresponding particles (see Table 1
below).\begin{center}
{\bf Table 1 : Particles and anti-particles}\\
\begin{tabular}{|c|c|c|c|c|}\hline
particle/ & electric & baryon & lepton & strange-\\
anti-particle & charge & number & number & ness  \\ \hline\hline
electron & -e & 0 &  1 & 0 \\ \hline
positron &  e & 0 & -1 & 0 \\ \hline\hline
proton   &  e & 1 &  0 & 0 \\ \hline
anti-proton & -e & -1 & 0 & 0 \\ \hline\hline
$K^+$   &  e & 0 & 0 &  1 \\ \hline
$K^-$   & -e & 0 & 0 & -1 \\ \hline
\end{tabular}
\end{center}
{\bf (c) Time Reversal - $T$}\\\\
Time reversal means reversing the direction of the time co-ordinate, i.e.,
$$t \ \stackrel{T}{\longrightarrow} \ -t$$
Symmetry of a physical system under time reversal simply means that all the
processes in the system are reversible.\\\\
Physical systems involving only strong and electromagnetic interactions are
symmetric under all three bilateral transformations listed above. But in
nature, other types of interactions, namely,  weak and gravitational
interactions exists. Weak interactions are known to violate $P$ as well as $T$.
We will discuss $P$ and $T$ violations later on in this article.\\\\
{\large{\bf CPT - theorem}}\\\\
If we take the whole universe as one system, which involves all four kinds of
interacions, then surely the system is not symmetric under $P$, $C$ and $T$
separately. In fact, the system violates all three symmetries. Then one may
ask, `` Is the system symmetric under certain combinations of these
transformations ? ", to which the answer turns out to be ``yes". Under the
combined operation of all three transformations physical laws remain unchanged.
More acurately, invariance under the combined action of $C$, $P$ and $T$ is a
consequence of relativistic invariance or {\it Lorentz invariance} of the law
of physics. Certain consequences of this theorem are :
\begin{itemize}
\item the mass of a particle and its anti-particle are exactly the same,
\item the total life-time, $\tau$, of an unstable particle and its anti-
particle are exactly the same,
\item the magnetic moment is equal and opposite for particle and anti-particle.
\end{itemize}
For example, mass of electron and positron are exactly same, the life-time of
$K^+$ is same as that of $K^-$ etc. Till date all the test for $CPT$ violation
has yielded in negative results.\\\\
{\large{\bf Parity violation}}\\\\
Before 1950's, parity was assumed to be a good symmetry of natural forces. But
in the early 50's the ``$\tau - \theta$ puzzle" posed a question on parity
symmetry. Two particles, then named as $\tau$ and $\theta$, were found to be
identical in almost every respect, such as their masses, life-times, charges,
spins, except their weak decay into pions,
$$\theta \ \longrightarrow \ \pi^+ \pi^0 \hspace{2cm} P=+1$$
$$\tau \ \longrightarrow \ \pi^+ \pi^+ \pi^-  \hspace{2cm} P=-1$$
Based on intrinsic parity of pions ($P_{\pi}=-1$), conservation of angular
momentum and conservation of parity, it was inferred that $\theta$ is even-
parity particle and $\tau$ is odd-parity particle. The puzzle was `` How can
two particles with otherwise identical kinematical properties have different
parities ? " This puzzle was solved by C. N. Yang and T. D. Lee, who proposed
that the weak interaction does not conserve parity and that the $\tau$ and
$\theta$ are the same particle, now renamed as $K^+$.
Yang and Lee then suggested experiments to search for parity violation, later
confirmed by C. S. Wu in $\beta$-decay of $^{60}Co$ nuclei. In the experiment
the $^{60}Co$ atoms were located in a thin surface layer of a single crystal of
Ce-Mg-nitrate, which was cooled to 0.003 K to reduce any thermal vibrations and
the whole system was placed in a strong magnetic field to align the nuclear
spin of $^{60}Co$ nuclei. If parity was a good symmetry then the out coming
$\beta$-particles should come in a symmetric way with respect to the spin
alignment of the nuclei. But it was observed that the $\beta$-particles are
emitted preferentially in a direction opposite to that of nuclear spin. Further
study of $\beta$-decay of $^{60}Co$ indicated that parity is not only violated,
but is violated maximally. Neutrinos produced in $\beta$-decay are found to be
left-handed only, and left-handed anti-neutrino was not observed. This
indicates that $P$ is violated along with violation of $C$ symmetry such that
$CP$ is conserved. These
properties of the weak interactions have been tested at very high precision and
at very low to very high energies for several decades. Nevertheless a certain
effect has only been recently experimentally observed (see Box 2) although it
was predicted soon after the work of Yang and Lee.\\\\
$CP$, first thought to be conserved in weak interactions, was later found to be
violated minutely in the neutral kaon system, and more
recently in the neutral B-meson system. We will discuss $CP$ violation and $T$
violation in the second part of this article.\\\\
%\newpage
\noindent
{\bf Box 1: History of Elementary Particle Physics}\\\\
It is very difficult to say exactly when elementary particle physics (EPP) came
into being. The first elementary particle to be discovered was the
{\em electron} by J.J. Thomson in 1897. Radioactivity was discovered in 1896 by
Becqurel, which is considered as the begining of nuclear physics and EPP was a
subset of it in early years. It might be fair to say that EPP separated from
nuclear physics around 50's.\\\\
Since the begining of the last century the experimentalist started probing
into the atom and in 1911 Rutherford found it to be composed of electrons and
positively charged {\it nucleus} and the nucleus of hydrogen atom was
identifiedas the {\it proton}. Then in 1932 Chadwick discovered the {\it
neutron}
and now the nucleus is understood to be a bound state of protons and neutrons.
Study of unstable nucleii hinted existence of massless chargeless particle,
called the {\it neutrino}, which was
postulated by Pauli in 1931 and experimentally observed by Reines and Cowan in
1956. The stability of nucleii against Coulomb repulsion was explained by
Yukawa in 1935 by introducing $\pi$-mesons as mediators of nuclear interaction
and which were first observed by Powell in 1947. After that a large number of
mesons were discovered. The existence of anti-particles was predicted by Dirac
in 1928 and the {\it positron}, the anti-particle of the electron was
discovered by Anderson in 1932. The heavier cousin of the electron, the muon
was discovered in Neddermeryer and Anderson in 1937.\\\\
By mid 50's a big list of {\it baryons}, the heavy particles, {\it mesons}, the
medium mass particles and {\it leptons}, the light particles, faced the
theorists which was then categorised based on their masses, spin, electric
charge, intrinsic parity etc. The quantum numbers of particles were found to
have a pattern in the categorised lists and hinted that all baryons and mesons
must have further constituents called {\it quarks}. The first quark model
talked about only three quarks, namely, up, down and strange but latter this
was extended to six quarks, where the sixth quark, the top quark
was discovered as late as in 1995. In late 60's weak interaction were
postulated to be caused by exchange of massive bosons, namely, $W^{\pm}$ and
$Z$ which were observed in collider experiments from 1983. Today observed
matter is understood to be made of six quarks, three leptons and three
neutrinos interacting among themselves via exchange of photons, three massive
bosons and eight gluons (carriers of strong interaction) along with one spin-0
massive boson called {\it higgs} which is not yet discovered.\\\\
{\bf Box 2:  The nuclear anapole moment}\\\\
Soon after the proposal that the weak interactions violate parity and its
experimental confirmation in the $^{60}Co$ system, Zel'dovich and Vaks proposed
that the weak interaction should induce an observable nuclear ``anapole"
moment, at a level that became possible to detect with technology that was
available only as recently as 1998.\\\\
The first term in the multipole expansion of the potential due to a electric
charge distribution, at a point outside the distribution is called the monopole
moment. The same term is called anapole moment when the potential is expanded
at a point inside the charge distribution. The anapole moment is zero if parity
is a good symmetry of the interaction among the charged particles. But in the
case of nuclei, the quarks interact among themselves via parity violating weak
interaction along with parity conserving strong and electromagnetic
interactions, and the anapole moment is non-zero and proportional to the
nuclear spin $I$. Experimentally it is very difficult to measure the nuclear
anapole moment (NAM) because it is caused by higher order weak interactions
among the quarks. Further, its contribution to parity non-conserving (PNC)
transition in atoms goes to zero when the exchanged photon is real. A very high
precision experiment (more than 1\% accuracy) on PNC transitions is required to
see the contribution from NAM. Other contributions to PNC transitions are
independent of the nuclear spin while NAM contributions depend on $I$. Thus to
measure the NAM one has to separate the $I$ dependent part from comparatively
large $I$ independent part, which became possible only as recently as 1998.\\\\
This example illustrates that while the electro-weak
interactions have been tested to very great precision at accelerator
experiments, there continue to be experimental challenges at the level of
table top experiments.\\\\ 
{\bf Suggested Readings}\\
A very inspiring book on the subject of symmetries is the following
classic:
\begin{itemize}
\item H. Weyl, {\it Symmetry}, Princeton University Press, Princeton, NJ, USA,
1952
\end{itemize}
Standard introductory reference to nuclear and elementary particle
physics are, e.g.,
\begin{itemize}
\item B. Povh, K. Rith, C. Scholz and F. Zetsch, 2nd edn.,
{\it Particles and Nuclei, An Introduction to the Physical Concepts},
Springer-Verlag, Berlin, 1999.
\item D. Griffiths, {\it Introduction to Elementary Particle Physics},
John Wiley \& Sons, New York, 1987
\item  G.D. Coughlan, J.E. Dodd {\it The Ideas of Particle Physics : An
Introduction for Scientists},Cambridge Univ. Pr., UK, 1993
\end{itemize}
The following articles in Resonance would provide the reader with
accounts on some of the subjects disccused here:
\begin{itemize}
\item Ashoke Sen, {\it Resonance}, Vol.5, No.1, p.4 2000
\item Rohini Godbole, {\it Resonance}, Vol.5, No.2, p.16 2000
\item Sourendu Gupta, {\it Resonance}, Vol.6, No.2, p.29 2001
\end{itemize}
A detailed treatment of the Cobalt experiment has already been made
available to the readers of Resonance in:
\begin{itemize}
\item Amit Roy, {\it Resonance}, Vol.6, No.8, p.32 2001
\end{itemize} 
\end{document}